# Anisotropic electronic structure and transport properties of the $\mathcal{H}$-0 hyperhoneycomb lattice

Marcos Veríssimo-Alves,[*] Rodrigo G. Amorim, and A. S. Martins

*Departamento de Física, ICEx,*

*Universidade Federal Fluminense*

*Volta Redonda, RJ, Brazil*

(Dated: May 31, 2016)

## Abstract

Carbon, being one of the most versatile elements of the periodic table, forms solids and molecules with often unusual properties. Recently, a novel family of three-dimensional graphitic carbon structures, the so-called hyperhoneycomb lattices, has been proposed, with the possibility of being topological insulators [1]. In this work, we present electronic structure calculations for one member ($\mathcal{H}$-0) of this family, using Density Functional Theory and non-equilibrium Green's functions transport calculations to show that the $\mathcal{H}$-0 structure should have strongly anisotropic electronic properties, being an insulator or a conductor depending on the crystalline orientation chosen for transport. Calculations in the framework of Extended Hückel Theory indicate that these properties can only be understood if one considers at least $2^{nd}$ nearest-neighbor interactions between carbon atoms, invalidating some of the conclusions of Ref. [1], at least for this particular material.

---

[*] Corresponding author. E-mail: marcos_verissimo@id.uff.br

## 1. INTRODUCTION

Graphene is one of the promising candidate for nanoelectronics, due to its linear electronic dispersion close to the Fermi level. This results in the so-called "massless Dirac Fermions" that make it not only a candidate for all sorts of devices, but as a laboratory for fundamental physics. In fact, experimental studies with graphene on fundamental physics include effects such as the observation of anomalous integer quantum Hall Effect at room temperature [2], and the Klein paradox [3] from quantum electrodynamics, to name just a few.

Nonetheless, materials based on carbon do appear in many allotropic forms of different and dimensionality, where it is the sole constituent, such as buckyballs (0D) [4], carbon nanotubes [5] and nanowires [6, 7] (1D), graphene (2D) [8] and diamond (3D) [9]. The electronic properties of each of these allotropes, in spite of being composed only of carbon, are completely different. Diamond, for example, is an insulator, and graphene is a zero-gap semiconductor; carbon nanotubes, on the other hand, can be either metallic or semiconducting, merely by a geometrical factor. This, in conjunction with their mechanical properties, confers carbon physics a great interest from the point of view of applications, with several devices based on graphene and carbon nanotubes having been proposed in the last few years [10].

Recently, inspired by recent experimental realizations of a three-dimensional spin-anisotropic harmonic honeycomb ($\mathcal{H}$-$n$ lattices) iridates [11], a family of new graphitic-like carbon networks was proposed by Mullen *et al* [1]. They study these so-called *hyper-honeycomb* networks by means of a simple nearest-neighbor one-$p$ orbital tight-binding Hamiltonian, which predicts quantized Hall conductivities in three dimensions for magnetic fields with toroidal geometry. Moreover, the presence of spin-orbit coupling would induce topological surface states, offering yet another laboratory for studying the spin Hall effect, or yet another application in devices based on the spin-torque transfer effect.

We have shown elsewhere [12] that these effects will not be present in such materials because the Hamiltonian of Ref. [1] does not take into account the orientation of $p$ orbitals adequately. In this article, we perform calculations in the framework of Density Functional Theory [13] and Extended Hückel Theory for the simplest member of the of the $\mathcal{H}$-$n$ family, the $\mathcal{H}$-0 lattice. We find that this lattice should have peculiar and interesting transport properties other than those proposed in Ref. [1], which could be of use in devices. We also show that, if these materials are to be studied within a tight-binding framework, 1$^{\text{st}}$ nearest-neighbor (n.n) interactions are insufficient to adequately describe the electronic structure and understand electronic transport properties; instead, at least 2$^{nd}$ n.n. interactions should be taken into account for a proper description of the physics and chemistry in this material.

## 2. METHODOLOGY

### 2.1. Density Functional Theory calculations

Density Functional Theory [13] calculations were carried out with the SIESTA package [14], using Troullier-Martins norm-conserving pseudopotentials [15] and the PBE exchange-correlation functional [16]. A 900 Ry mesh cutoff was used to ensure minimal "eggbox effect", and a 18 × 18 × 12 undisplaced Monkhorst-Pack grid [17] was used to sample the Brillouin Zone. A smearing temperature of 8 meV was used for the integrations in reciprocal space.

Atomic positions were fully relaxed using a conjugate-gradients algorithm, with a maximum force criterion of 0.01 eV/Å. The in-plane lattice parameter, $a$, and the out-of-plane to in-plane lattice parameter, $c/a$, were determined through Birch-Murnaghan fits [18] to sets of calculations where the $c/a$ ratio was determined for a given value of $a$ ; the global optimum value of $c/a$ was then determined through a final Murnaghan fit using the optimum $c/a$ for each value of $a$.

### 2.2. Extended Hückel Theory calculations

In our Extended Hückel Theory (EHT) calculations, a double-$\zeta$ basis set was employed to express the basis orbitals $\{\Phi_\nu\}$ as a sum of two Slater-Type Orbitals (STO). With this choice, in EHT the matrix elements of the Hamiltonian (**H**) and overlap (**S**) matrices can be expressed in terms of the atomic basis set as

$$\begin{aligned}
H_{\mu\mu} &= \langle \Phi_\mu | H | \Phi_\mu \rangle = E_{\mu\mu} \\
H_{\mu\nu} &= \frac{1}{2} K_{EHT} \left( H_{\mu\mu} + H_{\nu\nu} \right) S_{\mu\nu} \\
S_{\mu\nu} &= \langle \Phi_\mu | \Phi_\nu \rangle = \int \phi_\mu^* \phi_\nu d^3\mathbf{r},
\end{aligned} \quad (1)$$

where subscripts $\mu, \nu$ refer to basis orbitals and $K_{EHT}$ is an additional fitting parameter whose value is commonly set to 1.75 for molecules and 2.3 for solids [19]. In EHT, for each atom type, the on-site energies ($E_s$, $E_p$ and $E_d$), the $\zeta$ of the Slater Orbitals, and the first expansion coefficient $c_1$ need to be specified. The value of the second coefficient is then determined with the constraint of orbital normalization. The EHT band structure is obtained by solving the generalized eigenvalue problem

$$\mathbf{H}(\mathbf{k})\Psi_i(\mathbf{k}) = E_i(\mathbf{k})\mathbf{S}(\mathbf{k})\Psi_i(\mathbf{k}), \qquad (2)$$

where $\Psi_i(\mathbf{k})$ denotes the eigenvector of the $i^{th}$ band, and $\mathbf{k}$ is the Bloch wave vector within the first Brillouin Zone. The Hamiltonian and overlap matrices, $H(\mathbf{k})$ and $S(\mathbf{k})$, are calculated through

$$H_{i,j}(\mathbf{k}) = \sum_{j',m'} e^{i\mathbf{k}\cdot(R_{i0}-R_{j'm'})} H_{i0,j'm'} \qquad (3)$$

$$S_{i,j}(\mathbf{k}) = \sum_{j',m'} e^{i\mathbf{k}\cdot(R_{i0}-R_{j'm'})} S_{i0,j'm'}, \qquad (4)$$

where subscripts $i$ and $j$ label atoms within the unit cell and $m'$ is the unit cell index. The summation indices in equations 3 and 4 run over all atoms $j'$ in the unit cell $m'$ which are equivalent to atom $j$ in the reference unit cell ($m = 0$). The real-space matrix elements $H_{i0,j'm'}$ and $S_{i0,j'm'}$, constructed between atom $i$ in the reference unit cell and atom $j'$ in cell $m'$, are calculated through the Extended Hückel prescription, Eqs. (1).

### 2.3. Non-equilibrium Green's Functions transport calculations

For transport calculations, we have employed DFT combined with non-equilibrium Green's functions (NEGF), as implemented in the Transiesta code [20]. In this implementation, which follows the proposal by Caroli *et al.* [21], the device of interest is split into three segments, namely a scattering region and two semi-infinite electrodes to the left and right of the scattering region. The electrode regions are pristine bulk regions of the material, and for consistency, the charge density of the scattering and electrode regions were matched.

Since all transport calculations occur in the $z$ cartesian direction, we used an orthorhombic cell containing eight atoms and cell vectors coinciding with cartesian axes. Transport calculations along different directions thus involve different orientations of the unit cell, and bulk and scattering regions were oriented accordingly. Translational invariance is assumed along the $x$ and $y$ cartesian axes.

Using localized basis sets, it is possible to write the retarded Green's functions for the scattering region as

$$\mathcal{G}(E, V) = [E \times S_\text{S} - H_\text{S}[\rho] - \Sigma_\text{L}(E, V) - \Sigma_\text{R}(E, V)]^{-1} , \qquad (5)$$

where the overlap ($S_S$) and Hamiltonian ($H_S$) matrices are now calculated for the scattering region only. The semi-infinite leads are accounted for by means of the self-energies $\Sigma_{L/R}$, for left (L) and right (R) leads. For the system under consideration, the charge density is calculated self-consistently via Green's functions with an initial guess coming from DFT calculations. After convergence has been reached, the transmittance can be calculated as

$$T(E) = Tr\left[\Gamma_\text{L}(E, V)\mathcal{G}(E, V)\Gamma_\text{R}(E, V)\mathcal{G}^\dagger(E, V)\right], \qquad (6)$$

where the coupling matrices are given by $\Gamma_\alpha = i\left[\Sigma_\alpha - \Sigma_\alpha^\dagger\right]$, with $\alpha \equiv \{\text{L}, \text{R}\}$. The transmittance can be interpreted as the probability of an incoming electron with energy $E$ entering the scattering region from the left electrode and reaching the right electrode. The current density between two sites $M$ and $N$ is obtained with the Keldysh formalism and is given by

$$i(E)_{N \to M} = 4\frac{e}{h} \sum_{\substack{n \in N \\ m \in M}} \Im\left[\left\{\mathcal{G}(E)\Gamma_\text{L}\mathcal{G}^\dagger(E)\right\}_{mn} H_{nm}\right], \qquad (7)$$

where the sum is performed over all localized atomic orbitals $n$ and $m$ of the basis set, which are associated with sites $N$ and $M$, respectively. This quantity is called the "local current", and for zero bias calculation it is the projection of the transmittance between two given sites. For more details regarding the theory of electronic transport, we refer to the literature [22].

## 3. RESULTS

### 3.1. Geometry and band structure

The unit cell of the $\mathcal{H}$-0 lattice has four atoms, and belongs to the I4$_1$/amd space group. The unit cell, shown in Figure 1, has ideal lattice vectors $\vec{a}_1 = (1, 0, 0)$, $\vec{a}_2 = (0, 1, 0)$ and $\vec{a}_3 = (-0.5, 0.5, 1.732)$, in units of $a = 2.442$ Å, the cell parameter for graphene from our DFT calculations. In our setup, the basal plane is parallel to the cartesian $xy$ plane. The $\mathcal{H}$-0 structure is formed by sets of zigzag chains, one being formed by atom pair $1-2$ and the other by atom pair $3-4$. In our setup, the chain formed by atoms $1-2$ is parallel to the $xz$ plane, while the chain formed by atoms $3-4$ is parallel to the yz plane. These two sets of chains are rotated by 90° relative to one another, and are connected by a bond lying along the direction perpendicular to the basal plane of the unit cell.

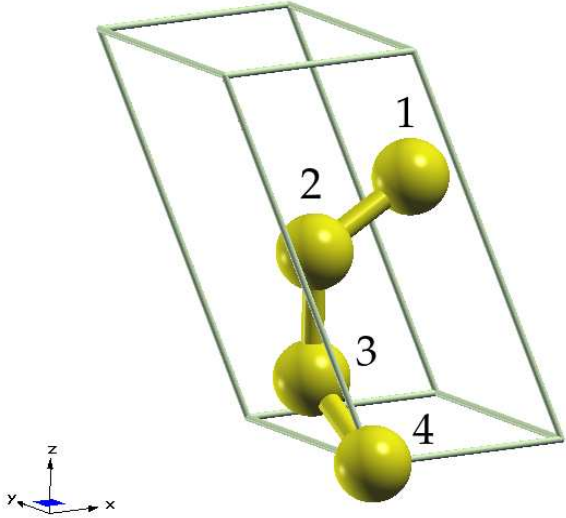

FIG. 1. The unit cell for the hyperhoneycomb $\mathcal{H}$-0 structure. Atoms have the same numbering as in Ref. [1]. Atoms 1 and 2 form a zigzag chain parallel to the $xz$ plane and atoms 3 and 4 form a zigzag chain parallel to the $yz$ plane. Due to periodic boundary conditions, atoms 1 and 4 are nearest neighbors. Intra-chain bonds, due to $sp^2$-like hybridization, have both a $\sigma$ component parallel to the chain and a $\pi$ component due to unhybridized, out-of-plane ($\perp$) $p$ orbitals perpendicular to the chains.

Relaxation of atomic geometry at the ideal graphene lattice parameter ($a_0^{graphene} = 2.44$ Å) reveals two different bond lengths: a shorter bond length, for C-C bonds within a given zigzag chain, and a longer one, for C-C bonds which connect chains along the direction perpendicular to the unit cell's basal plane. In the remainder of the text, we shall call the former kind of bond *intra-chain* bond, and the latter, *inter-chain* bond.

After full geometry optimization, we find that the optimized unit cell and atomic geometries differ significantly from their ideal values. The lattice parameter is now $a = 2.553$ Å, and the lattice vectors are $\vec{a}_1 = (1, 0, 0)$, $\vec{a}_2 = (0, 1, 0)$ and $\vec{a}_3 = (-0.5, 0.5, 1.705)$, in units of $a$. The optimized intra- and inter-chain C-C bond lengths are $a_{C-C}^{intra} = 1.445$ Å and $a_{C-C}^{inter} = 1.499$ Å respectively, and the angle between three consecutive atoms in a zigzag chain is $\theta_{intra} = 124.1°$, indicating a deviation from ideal $sp^2$ hybridization.

The rationale for different intra-chain and inter-chain bond lengths can be established as follows. We first note that, in a tight-binding picture with $2s$ and $2p$ orbitals included, for a given C atom two of the $p$ orbitals will be linearly combined with the atom's $s$-orbital to form three $sp^2$ hybrids. These hybrids, contained in the same plane and forming 120° with each other, will form strong $\sigma$-bonds with other $sp^2$ orbitals from neighboring C atoms and in graphene they form the well-known honeycomb lattice. The remaining, non-hybridized $p$ orbital will be oriented perpendicular to the plane formed by graphene C atoms and is usually referred to by $p_z$, but in this work we will refer to them by out-of-plane $p$-orbitals, or $p_\perp$ for short, for reasons that will soon become clear. These orbitals will form weaker $\pi$-bonds with the neighboring $p_\perp$ orbitals, being the ones responsible for electronic conduction along the graphene plane.

Therefore, in the structure shown in Fig. 1, the chain formed by atoms 1 and 2 will have $p_\perp$ orbitals perpendicular to the $xz$ plane, while the chain formed by atoms 3 and 4 will have $p_\perp$ orbitals perpendicular to the $yz$ plane. Each individual chain has an extended one-dimensional $\pi$-bond network parallel to the plane in which it is contained, as well as $\sigma$ bonds connecting the atoms in the chains. However, now the overlap between $p_\perp$ orbitals of chains of atoms $1-2$ and $3-4$ are also perpendicular to each other, and their overlap is null. Also, each $p_\perp$ now only forms two $\pi$ bonds, in contrast to the three it would form in graphene.

While the 90° rotation between zigzag chains still allows for the formation of a $\sigma$-bond through the $sp^2$ hybrids that lie along the line that connects neighboring perpendicular chains, no $\pi$-bonds are formed between them now, and we propose that this lack of a $\pi$ bond weakens the inter-chain C-C bonds, making them longer than their intra-chain counterparts. In order to test this hypothesis, we resort to a simple model of an ethylene ($C_2H_4$) molecule, in which each C atom has three hybrid $sp^2$ orbitals. One $sp^2$ hybrid will form a $\sigma$-bond with another $sp^2$ orbital from the other C atom and, on each C atom, the other two $sp^2$ hybrids will bond to the H atoms. The remaining $p_\perp$ orbitals will form a $\pi$-bond, as shown in Figure 2 (a).

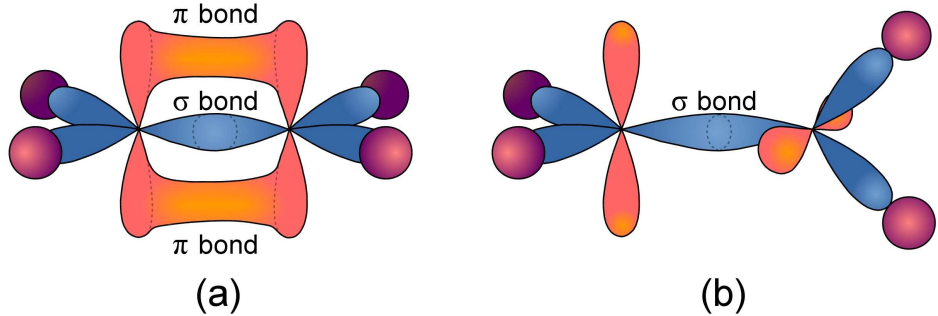

FIG. 2. (a) Ethylene molecule, $C_2H_4$, in its ground state. $sp^2$ hybrids from one $CH_2$ unit will form one $\sigma$ bond with the C atom from the other $CH_2$ unit and non-hybridized $p_\perp$ orbitals will form a $\pi$ bond. The remaining $sp^2$ hybrids bond to H atoms. (b) Rotating the plane of $sp^2$ hybrids of one $CH_2$ unit around the $\sigma$ bond weakens it due to breaking of the $\pi$ bond.

We test our hypothesis by performing two simple DFT calculations for the ethylene molecule: one for its equilibrium geometry, where all $sp^2$ hybrids are contained in the same plane, and another where *pi*-bonds are broken by rotation of one of the $CH_2$ units by 90°. After full relaxation of the two geometries, we find that the C-C bond for the twisted geometry is 0.085 Å larger than the C-C bond length in the equilibrium geometry. This is comparable to the 0.054 Å difference in lengths of the inter- and intra-chain C-C bond lengths in the $\mathcal{H}$-0 lattice. Hence, we propose that this is the basic mechanism behind the different inter- and intra-chain bond lengths in the $\mathcal{H}$-0 lattice, with the discrepancies in the bond length differences being due to the $\mathcal{H}$-0 being a solid.

We now turn to the electronic structure of the $\mathcal{H}$-0 lattice. Figure 3 shows the band structure from our DFT calculations and those from the TB calculations of Ref. [1]. We can see that while both DFT and the one-$p_\perp$ orbital approximations predict the $\mathcal{H}$-0 structure to be metallic, the band structure is completely different in the vicinity of the Fermi level, $E_F$. In particular, the particle-hole symmetry from the one-$p_\perp$ orbital band structure is completely lost when one fully takes into account the electronic structure. As we have shown elsewhere, this is due to the fact that in setting up the TB Hamiltonian in Ref. [1] the carbon chemistry is incorrectly taken into account: the TB band structure shown in Fig. 3-(b) is, in fact, an $s$-orbital band structure, and not a $p$-orbital one. As we will show now, a nearest-neighbor (n.n.) TB Hamiltonian, even if taking into account the correct orientation of $p$ orbitals, is insufficient to describe the electronic structure of the $\mathcal{H}$-0 lattice.

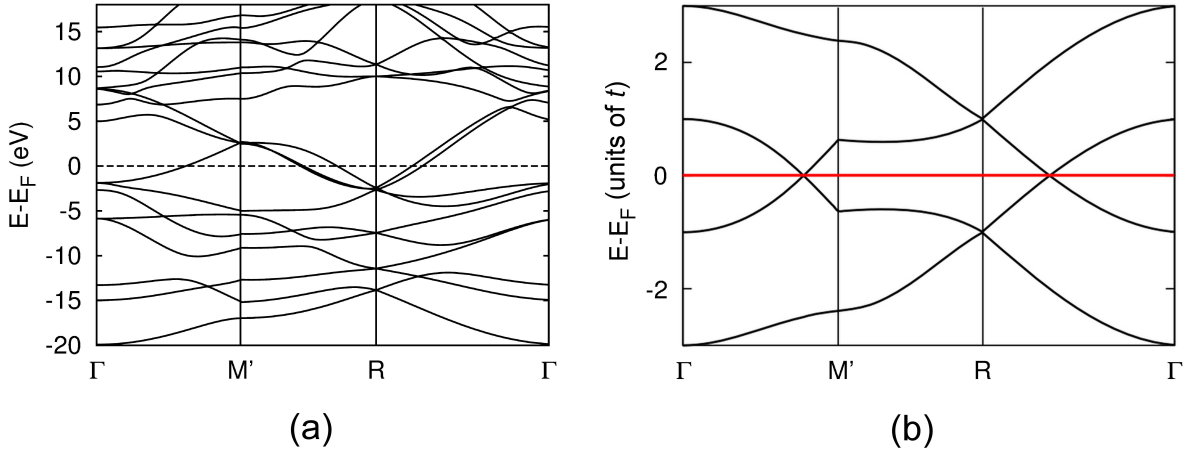

FIG. 3. Band structure from (a) DFT and (b) from the one-$p_\perp$ orbital TB calculation of Ref. [1] for the $\mathcal{H}$-0 structure. The Fermi level is set to zero in either case.

While DFT calculations provide a fully detailed description of the electronic structure, they do so at the expense of simplicity of description. It is harder to make controlled approximations in DFT calculations than in other theoretical frameworks, such as TB theory. In order to try and explain the discrepancies between DFT and the TB calculations of Ref. [1], we resort to the Extended Hückel Theory [23], which is itself also a TB approximation. Figure 4 shows the band structures resulting from an EHT calculation for $1^{st}$, $2^{nd}$ and $4^{th}$ nearest-neighbor (n.n.) hoppings, using a parametrization for graphene [24]. From these band structures it is clear that it is mandatory to go beyond the simple nearest-neighbor hopping approximation if one wants to reproduce the DFT bands: a $1^{st}$ n.n. Hückel approximation yields a band structure completely discrepant from those predicted by DFT. It is especially problematic that flat bands appear along one of the high symmetry lines in our $1^{st}$ n.n. Hückel calculation exactly on top of the Fermi level. These flat bands indicate no significant overlap between $p$ orbitals, which confer these bands' predominant character, occurs. Since these flat bands disappear when considering hoppings up to $2^{nd}$ n.n. and increasingly good agreement between the DFT and Hückel calculations is achieved as interactions between farther neighbors are considered, our EHT calculations show that only $1^{st}$ n.n. interactions, as considered in Ref. [1], are insufficient to describe the $\mathcal{H}$-0 lattice's electronic structure. This will be important for the interpretation of the transport properties of the $\mathcal{H}$-0 lattice, which we will analyze next.

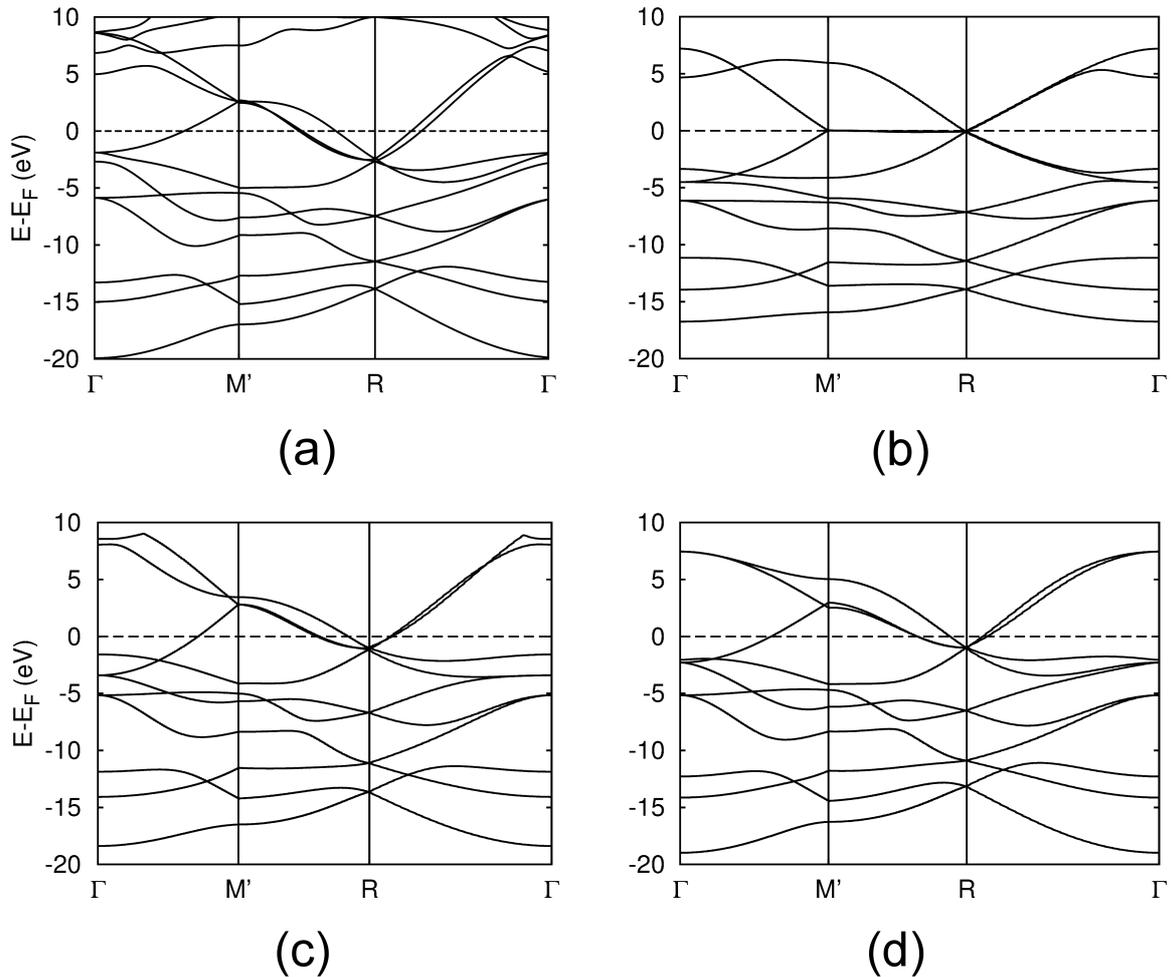

FIG. 4. Band structures for the $\mathcal{H}$-0 structure. (a) Band structure from DFT calculation. (b)-(d) Band structures from EHT calculations for $1^{st}$, $2^{nd}$ and $4^{th}$ nearest-neighbor hopping, respectively.

### 3.2. Interplay of atomic geometry, electronic structure and transport properties

In this section, we proceed to a further analysis of the interplay of the atomic geometry of the $\mathcal{H}$-0 lattice and its electronic structure, and how they influence the transport properties of this material. For ease of analysis, we now define the nomenclature we will use to refer to the two other $p$ orbitals perpendicular to $p_\perp$ orbitals. Generally speaking, we will have one
10

$p$ orbital parallel to the inter-chain bonds, to which we will refer to $p_{ic}$, and another which will always be parallel to the plane of the zigzag chain to which it belongs, which we shall denote as $p_{zp}$. For the zigzag chains formed by atom pair 1-2 in Fig. 1, for example, $p_{zp}$ would lie in the $xz$ plane, while for the zigzag chain formed by atom pair 3-4 it would lie in the $yz$ plane. $p_{ic}$ orbitals, in turn, would lie along the bonds between atoms 1-4 and atoms 2-3, parallel to the cartesian $z$-axis.

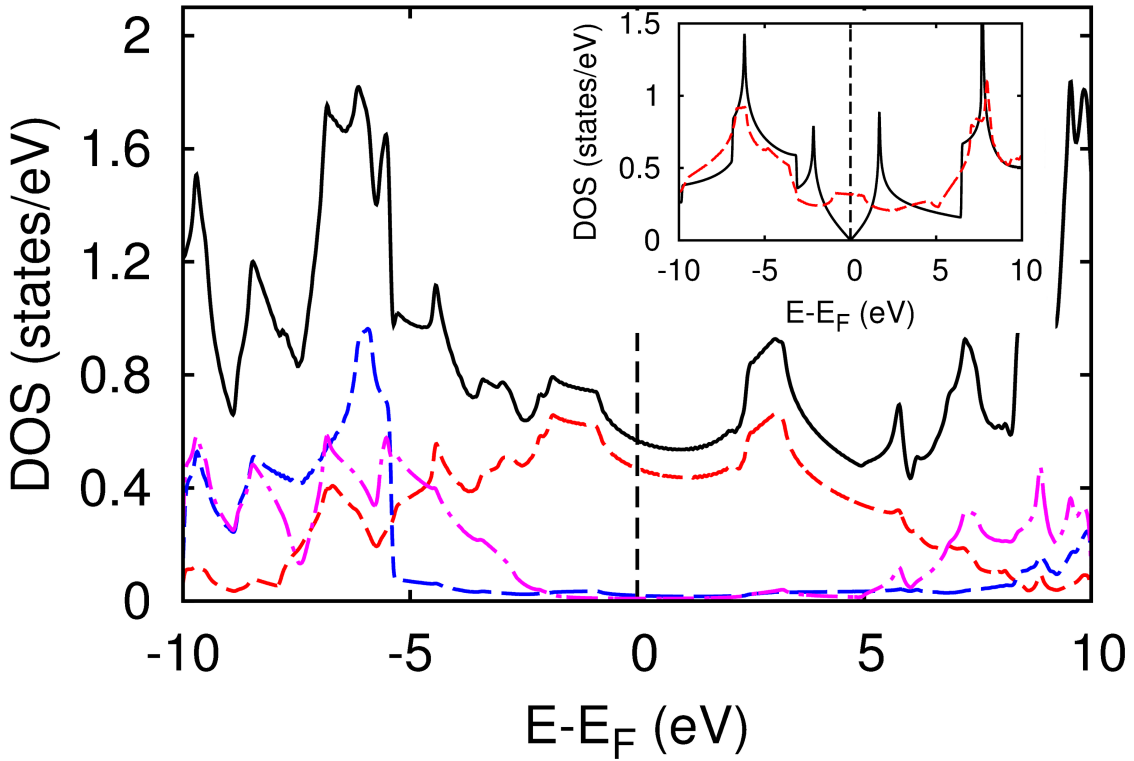

FIG. 5. DFT DOS for the $\mathcal{H}$-0 lattice. Black, solid line: total DOS; red, dashed line: PDOS from $p_\perp$ orbitals; blue, dashed line: PDOS for $p_{zp}$ orbitals (perpendicular to $p_\perp$, parallel to the zigzag chain that contains it); magenta, long dash-dotted line: PDOS for $p_{ic}$ orbitals (parallel to inter-chain bonds, also perpendicular to $p_\perp$). Inset: DOS for an isolated graphene sheet (black lines) and for an infinite stack of graphene sheets (red, dashed lines), where the separation between layers is the $\mathcal{H}$-0 structure's lattice parameter, $a = 2.553$ Å.

Figure 5 shows the electronic density of states (DOS) for the relaxed $\mathcal{H}$-0 lattice, along with the Projected DOS (PDOS) for all $p$ orbitals. The portion of the DOS in the vicinity of $E_F$ is largely due to $p_\perp$ orbitals, as expected, but it displays two curious flat features, one just below and one just above $E_F$. In order to understand the origin of the flat features close to $E_F$, we perform a simple DOS calculation for an essentially isolated graphene sheet, and an infinite AA stacking of graphene sheets separated by the $\mathcal{H}$-0 lattice parameter, $a$.

The inset of Figure 5 shows the DOS for both an isolated graphene sheet and for an infinite stacking of graphene sheets. For the stacked sheets, a flat feature appears at $E_F$, as a result of the overlap of the $\pi$-bond networks of adjacent graphene sheets, leading to the formation of $\sigma$ bonds by $p_\perp$ orbitals of adjacent graphene sheets. In this geometry, besides interacting with the $p_\perp$ orbital immediately above it, each $p_\perp$ orbital interacts with three other $p_\perp$ orbitals present in the plane of the graphene sheet.

In the $\mathcal{H}$-0 structure, however, the flat feature is displaced towards lower energies, lying below $E_F$, while the DOS still has a sizable value at this energy. Our explanation is that the flat feature below $E_F$ is due mainly to the interaction between $p_\perp$ in adjacent parallel zigzag chains, while the finite DOS at $E_F$ is due mainly to the 1-D $\pi$ bond network of individual zigzag chains. This explanation, as we will see later, is consistent with the transport properties predicted for the $\mathcal{H}$-0 lattice.

From Fig. 5, it is also evident that $p_{zp}$ and $p_{ic}$ orbitals will not contribute to electronic transport at biases lower than 0.5 V, since their contribution to the total DOS only becomes non-negligible at 1.5 eV below $E_F$. Considering the PDOS and the geometry of the $\mathcal{H}$-0 structure therefore suggests it might have strongly anisotropic electronic transport properties, which we now analyze in detail through non-equilibrium Green's functions calculations. Here, we consider two possible electronic transport directions: one along either of the zigzag chains in the unit cell, and one along the inter-chain direction. As explained in Section 2.3, electronic transport is assumed to occur along the cartesian $z$ direction. The two different scattering regions used for our calculations, which involve re-orienting the unit cell accordingly, are shown in Figure 6 (b) (inter-chain bonds along cartesian $z$ direction) and 4 (c)-(e) (zigzag chain along the cartesian $z$ direction). Only one electronic transport calculation is performed for the zigzag chains, since the $\mathcal{H}$-0 structure has a 90° rotation symmetry around the inter-chain bond direction.

Figure 6 (a) shows the transmission curves for the two scattering regions considered. For electronic transport along the inter-chain direction, transmission in the vicinity of Fermi level (black) is indeed zero, whereas for the scattering region with zigzag chains parallel to the transport direction, twelve electronic channels are present, thus confirming our hypothesis of strongly anisotropic electronic transport. We should stress that the local current we present is obtained for zero bias calculations. It is, strictly speaking, the projection of the transmittance between two states, and not a real electronic current. Nevertheless, for weak electronic bias, the real electronic current should not be too different from the presented results.

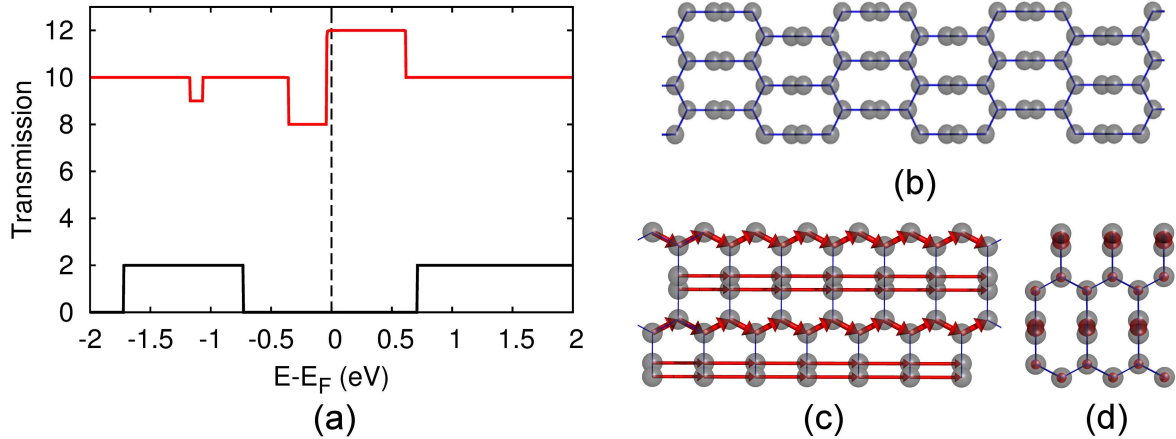

FIG. 6. (a) Transmission curves for electronic transport along the direction parallel to inter-chain chemical bonds (black line) and along zigzag chains (red line). (b)-(d) Total current flowing through the $\mathcal{H}$-0 structure in different orientations. Electron flow is indicated by red arrows, and its intensity is proportional to arrow thickness. (b) Total current for inter-chain bonds parallel to the electronic transport direction. No electronic current is observed. (c) Side view of the structure. Electron flow happens along zigzag chains, with a smaller contribution from tunneling between the perpendicular zigzag chains. (d) Rear view of the $\mathcal{H}$-0 orientation of Fig. 6 (d). Current flow is perpendicular to the page.

Figures 6 (b)-(e) show the local current for both scattering region orientations, with electronic flow happening in the direction indicated by the arrows, and arrow thickness proportional to current intensity (projected transmittance). Clearly in the case where inter-chain bonds lie along the transport direction, no current flows through the structure, and along this direction, the $\mathcal{H}$-0 lattice is an insulator. On the other hand, when zigzag chains lie along the transport direction, we can separate the current into two components with different intensities: a more intense current flow along the zigzag chains lying parallel to the electronic transport direction, and a less intense one between the zigzag chains perpendicular to those where more intense current flows.

Both currents can be understood with the aid of the PDOS shown in Figure 5 and the EHT calculations presented in Section 3.1. The more intense current, which occurs along the zigzag chains, takes place along the $\pi$ bond network formed by the $p_\perp$ orbitals. Therefore the energy levels corresponding to the $\pi$-bond network should be closer to $E_F$. The less intense current is due to tunneling between the zigzag chains perpendicular to the transport direction and therefore occurs through overlapping $p_\perp$ orbitals from adjacent parallel chains. In a TB interpretation, this is a $2^{nd}$ n.n. interaction between $p$ orbitals, arising from the proximity of the zigzag chains; in turn the proximity of $p_\perp$ orbitals is related, as we conjectured early, to the hump in the PDOS at $E_F$ shown in the inset of Fig. 5. Once more, interactions between $2^{nd}$ n.n. lead to important and interesting effects, this time in the electronic transport properties of the lattice, which would not be captured by a simple $1^{st}$ n.n. TB approximation, such as that used in Ref. [1].

In order to understand why transport is negligible along the inter-chain directions, we resort to a COHP [26] analysis to determine the bonding/anti-bonding character of the intra- and inter-chain chemical bonds. Figures 7 (a) and 7 (b) show the COHP curves for the orbitals involved in the bonding for the two different chemical bonds present in the $\mathcal{H}$-0 lattice. The COHP curves reveal that in a zigzag chain the chemical bond has a greater bonding character especially due to the $\pi$-bonds formed by the $p_\perp$ orbitals of the C atoms. However, COHP for inter-chain C-C bonds reveals a non-negligible anti-bonding character close to $E_F$, consistent with the fact that transport along the inter-chain direction is null.

The transport properties predicted for the $\mathcal{H}$-0 lattice could allow for its use in two applications: gas sensors and a nanodynamometer. For use as a gas sensor, a thin film of the material, grown in such that a surface in the orientation of Fig. 6-b is present, might show differences in measured electronic surface currents. In Fig. 6-c, consider the uppermost zigzag chains as the surface. Electronic transport should happen both along the zigzag C chain, on the $\pi$-bond network formed by $p_\perp$ orbitals, or along the perpendicular direction, due to frontal overlap of $p_\perp$ orbitals of adjacent zigzag C chains on the surface. It is likely that adsorbed molecules would act as scattering centers to the electronic currents due to changes in hybridization of $p_{zp}$ and, possibly, of $p_\perp$ orbitals. As in graphene-based gas sensors [27], each individual molecule will induce a particular hybridization change, leading to sensitivity variation.

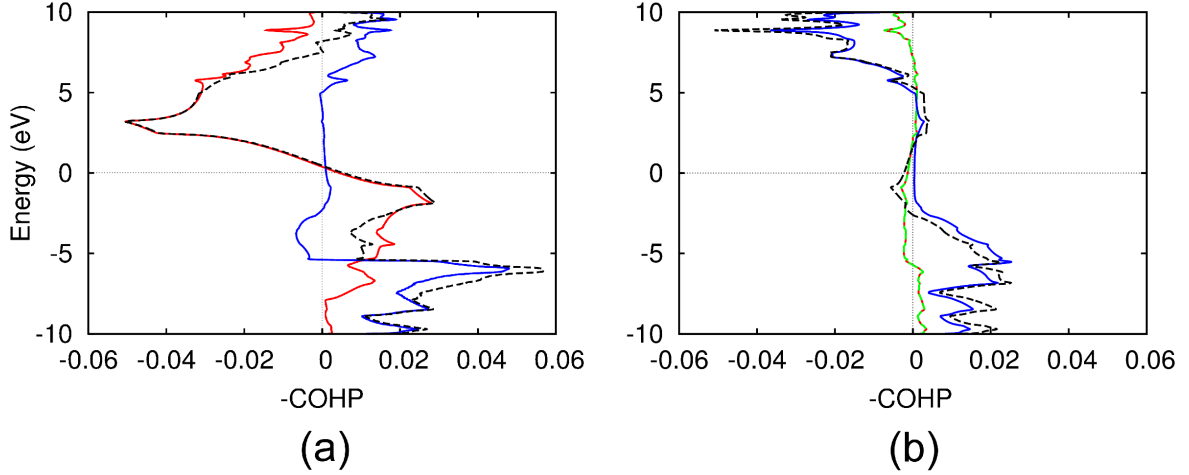

FIG. 7. (a) COHP curve for $p$ orbitals taking part in the intra-chain chemical bonds (atoms 1-2 or 3-4 of Fig. 1). Red, solid line: $p_\perp - p_\perp$ orbitals; blue, solid line: COHP for all $p_{zp}, p_{ic}$ orbitals; black, dashed line: COHP for all $p$ orbitals. In both cases, only non-zero COHP curves are displayed. (b) COHP curve for $p$ orbitals taking part in the inter-chain chemical bonds (atoms 2-3 or 1-4 of Fig. 1). Subscripts $ch1$ and $ch2$ refer to each of the zigzag chains in the unit cell. Red, solid line: $p_{\perp,ch1} - p_{zp,ch2}$ orbitals; green, dashed line: $p_{zp,ch1} - p_{\perp,ch2}$ orbitals; blue, solid line: COHP for $p_{ic,ch1} - p_{ic,ch2}$ orbitals; black, dashed line: COHP for all $p$ orbitals.

For use as a nanodynamometer, the frontal overlap of $p_\perp$ orbitals would be the key player. A force applied perpendicular to sets of parallel zigzag chains would deform the material, thus leading to a change in the overlap of $p_\perp$ orbitals which, in turn, would cause the electronic currents to vary. The current differences would be therefore directly linked to the deformation of the material, allowing for determination of the applied force after calibration. The principle is similar to the nanodynamometer of Poklonski *et al* [28], who propose such a device based on bilayer graphene, but it could be more practical than the one made of graphene bilayers since with the $\mathcal{H}$-0 lattice there would be no possibility of slipping layers, in contrast with the bilayer graphene device.

## 4. CONCLUSIONS

In this work, the structural, electronic and transport properties of carbon-based graphitic family materials ($\mathcal{H}$-0), as proposed in Ref. [1], were discussed using DFT, EHT and non-equilibrium Green's functions transport calculations. We have shown that its geometry implies a spatial orientation of unhybridized $p_\perp$ orbitals (usually denoted as $p_z$) such that it has 1D $\pi$ bond networks, in contrast with the 2D network of graphene. This confers it anisotropic electronic transport properties, being an insulator or a conductor depending on the crystalline direction considered.

We have also shown that a simple tight-binding model with only $1^{st}$ n.n. interactions and one $p$ orbital per C atom is unable to adequately describe the electronic structure and transport properties of this material. In particular, the electronic transport properties of the $\mathcal{H}$-0 can only be accounted for if $2^{nd}$ n.n. interactions are considered, since they are related to the $\sigma$ overlap between $p_\perp$ orbitals from adjacent parallel zigzag chains of C atoms in the solid. Our calculations indicate that, while this $\sigma$ interaction is one of the interactions that causes the $\mathcal{H}$-0 lattice to be metallic, electronic transport is due mostly to the 1-D $\pi$ bond network of individual zigzag chains. We have also discussed possible applications of this material in devices such as gas sensors and nanodynamometers; further calculations to assess the usefulness of the $\mathcal{H}$-0 lattice in such devices are currently under way.

## ACKNOWLEDGMENTS

M. Veríssimo-Alves and A. S. Martins acknowledge funding agency FAPERJ for financial support through grants E-26/111.397/2014 and E-26/112.554/2012.

---